\documentclass[]{article}
\usepackage{amsmath}
\usepackage{amsfonts}
\usepackage{amssymb}
\usepackage{graphicx}
\usepackage{amsfonts}
\usepackage[english]{babel}
\usepackage[utf8]{inputenc}
\usepackage{times}
\usepackage[T1]{fontenc}

\begin{document}

\title{Wheeler-DeWitt Equation for 4D Supermetric and ADM with Massless Scalar Field as Internal Time}
\author{Leonid Perlov\\
Department of Physics, University of Massachusetts,  Boston\\
leonid.perlov@umb.edu
}
\date{ January 7, 2015}

\maketitle

\begin{abstract}
The main result of this paper is the 4-dimensional supermetric version of the Wheeler-DeWitt equation, that uses only one time variable for the both roles - as internal time and for the ADM split, as Hamiltonian evolution parameter. We study the ADM split with respect to the scalar massless field serving as internal time. The 4-dimensional hyper-surfaces $\Sigma_{\phi = const}$ span the 5-dimensional space with the scalar field being the fifth coordinate. As a result we obtain the analog of the Wheeler-DeWitt equation for the 4-dimensional supermetric. We compare the ADM action with the non-compactified Kaluza-Klein action for the same physical space and obtain the equation for the extrinsic curvature and the scalar massless field. 
\end{abstract}

\section{Introduction}

The massless scalar field serves as internal time in Einstein and Wheeler-DeWitt equations used in Loop Quantum Cosmology $\cite{AshtekarBojowald} \cite{AshtekarSingh}$. One regards the massless scalar field  as a parameter connecting the space-time coordinates. By doing that one does not replace the Lorentz time variable with the massless scalar field, but instead adds one more time variable, since the Lorentz time plays an important role in WDW derivation being an ADM split variable. The original form of the equation  $\cite{DeWitt}$  uses the ADM split of the 4-dimensional Riemann space by the $\Sigma_{t = const}$ hyper-surfaces. Therefore there are eventually two time variables in the Loop Quantum Cosmology version of WDW equation.

The goal of the current research is to use only one time variable -  the massless scalar field, for both purposes: the ADM split and as internal time. In order to achieve this we consider a 4-dimensional Lorentz space as embedded into the 5-dimensional space by adding the massless scalar field as a fifth coordinate. We define a new ADM split  by the  $\Sigma_{\phi = const}$, scalar field equals constant hyper-surfaces, and derive the analog of the Wheeler-DeWitt equation for the 4-dimensional supermetric. Such equation will describe the evolution of the 4-dimensional Lorentz hyper-surfaces along the scalar massless field. 

We derive the dynamics equation by considering the Hilbert-Einstein 5-dimensional action in two different forms: the new ADM split and the non-compactified version of the Kaluza-Klein formalism. 

The Kaluza-Klein 5-dimensional unified theory was known for a long time $\cite{Kaluza}$. By varying the Hilbert-Einstein vacuum action in the 5-dimensional space, one obtains the 4-dimensional Einstein equation with the electromagnetic and matter stress-tensor on the right hand side plus the Maxwell equation. The magic is probably due to the Campbell's conjecture stating that any 4-dimensional Riemann space can be embedded into the 5-dimensional Ricci flat space $\cite{PWessonBook}$. One had yet to explain why the fifth dimension was not observable in the classical world. To address it, Klein introduced the compactified model, $\cite{Klein}$, where the fifth coordinate belonged to a compact group. 

In the WDW equation used in the Loop Quantum Cosmology the fifth coordinate plays the role of internal time and has a physical meaning of the scalar massless field. Therefore there is no need in any compactification procedures. Hence we use in this paper the non-compactified Kaluza-Klein formalism.

A few recent papers studied the conventional ADM in the 5-dimensional space $\cite{Montani}$ $\cite{PoncedeLeon}$, where the split was carried out by the time-constant $\Sigma_{t=const}$  hyper-surfaces, rather than by matter-constant $\Sigma_{\phi = const}$  hyper-surfaces used in this paper. Following that research, the  3-dimensional supermetric Wheeler-DeWitt equation was studied in  $\cite{Pavsic}$.

The paper is organized as follows. In section $\ref{sec:5D ADM Split}$, we carry out the new ADM split by the $\Sigma_{\phi = const}$ hyper-surfaces. In section $\ref{sec:Kaluza-Kliein}$, we consider Kaluza-Klein non-compactified representation. In section $\ref{sec:Wheeler-DeWitt}$, we derive the 4-dimensional supermetric Wheeler-DeWitt equation. We conclude with the discussion in the final section $\ref{sec:Discussion}$.\\ 

\section[test]{ADM Split with the Massless Scalar Field as Internal Time}
\label{sec:5D ADM Split}

\subsection{5D ADM Spacelike and Timelike Split}
\label{subsec:5dadmsplit}

We begin with the ADM split in the 5-dimensional Riemann space with respect to the scalar massless field as internal time. The Riemann space coordinates are $(x_0, x_1, x_2, x_3, y)$, the metric below is a 4-dimensional Lorentz metric plus the massless scalar field corresponding to the fifth coordinate. For the 5-dimensional metric we use the notation $\tilde{g}_{AB}$ for the 4-dimensional metric we use $g_{\mu \nu}$ ,  where the indeces are $ \mu , \nu = 0, \mbox{...}, 3$  and  $ A, B = 0, \mbox{...}, 4 $\\[2ex]
$
\begin{pmatrix}
g_{\mu \nu} &  \cdots  & 0 \\
\vdots  & \ddots & \vdots \\
0 & 0  &  \epsilon {\phi}^2 \\
\end{pmatrix}
$\\[2ex]

$ g_{\mu \nu}$ - is a four dimensional Lorentz space metric, $\phi(x_{\mu})$ is a massless scalar field, $\epsilon$ equals to 1 for the spacelike and -1 for the timelike split hypersurface cases.

Consider the 4 + 1 split of the 5-dimensional space-time-matter manifold $M$ by the four dimensional hyper-surfaces $\Sigma_{\phi = const}$.\\[2ex]

\begin{figure}[htbp]
\centering
\includegraphics[width = 100mm, height=60mm]{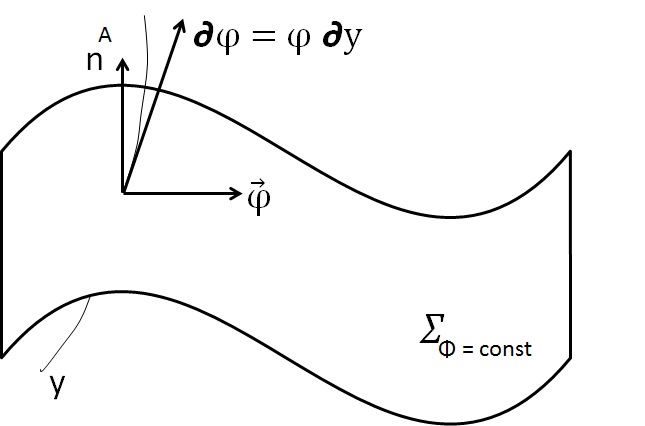}
\caption{ The 4+1 split with the lapse function $\phi$ and the shift vector $\vec{\phi}$} 
\end{figure}

We proceed similar to 3+1 case with $\phi$ playing the role of  the time lapse N, and $\vec{\phi}$ playing the role of the time shift $\vec{N}$, with the very important difference: we consider both $\Sigma_{\phi = const}$ spacelike and timelike cases. The spacelike case corresponds to $\epsilon = 1$, while the timelike to $\epsilon = -1$.

We project the 5-dimensional covariant derivative and decompose it into lapse and shift:

\begin{equation}
\label{5dcovariantderivative}
^5\nabla_{u}v = \epsilon \tilde{g} (^5\nabla_{u}v, n)n + (^5\nabla_{u}v  -  \epsilon \tilde{g} (^5\nabla_{u}v, n)n )
\end{equation}

\begin {equation}
\tilde{g}({n}, {n}) = \epsilon ,  \mbox{and} \;  \tilde{g}({n}, {v})  = 0 , \forall \; {v} \in T_p \Sigma
\end{equation}

We then define the extrinsic curvature to be the first term in  ($\ref{5dcovariantderivative}$) \\
\begin{equation}
K(u, v) n =  \epsilon \tilde{g} (^5\nabla_{u}v, n)n = -\epsilon \tilde{g} (^5\nabla_{u}n, v)n
\end{equation}

As for the second term in  ($\ref{5dcovariantderivative}$), it is a 4-dimensional covariant derivative in the 4-dimensional space. 

\begin{equation}
^4\nabla_u v = ^5\nabla_u v - \epsilon \tilde{g} (^5\nabla_{u}v, n)n = ^5\nabla_u v - K(u, v)n
\end{equation} 

It is easy to prove that the expression above is a 4-dimensional connection (covariant derivative).  One should prove that it satisfies the Leibniz law,  preserves the metric and is torsion free.

By applying the above formula to the massless scalar field we obtain (see Figure 1)

\begin{equation}
\partial_y = \epsilon \phi n + \vec{\phi} \qquad \phi = \tilde{g}(\partial_y, n)
\end{equation}

\begin{equation}
\label{5dnablacoord}
^5\nabla_{\mu}\partial_{\nu} = \epsilon K_{\mu \nu} n + ^4\Gamma^{\lambda}_{\mu \nu} \partial_{\lambda}
\end{equation}

\begin{equation}
\label{5dnablan}
^5\nabla_{\mu} n = - \epsilon K^{\lambda}_{\mu} \partial_{\lambda}
\end{equation}
 
We use ($\ref{5dnablan}$) in order to project the 5-dimensional Riemann curvature tensor to the 4-dimensional subspace and obtain the Gauss and Codazzi equations for the 5-dimensional space.

By the Riemann tensor definition:
\begin{equation}
\label{riemanntensor}
^5R (\partial_{\mu}, \partial_{\nu})\partial_{\lambda} = {^5\nabla}_{\nu}{^5\nabla}_{\mu}\partial_{\lambda} - {^5\nabla}_{\mu}{^5\nabla}_{\nu}\partial_{\lambda}
\end{equation}

and ($\ref{5dnablacoord}$), ($\ref{5dnablan}$)  we obtain the Gauss equation in 5-dimensional spacelike ADM

\begin{equation}
\label{5dgauss}
^5R^4_{\mu \nu \lambda} = {^4\nabla}_{\mu}K_{\nu \lambda} - {^4\nabla}_{\nu}K_{\mu \lambda}
\end{equation}

and Codazzi equation for the 5-dimensional ADM

\begin{equation}
\label{5dcodazzi}
{^5R}^{\sigma}_{\mu \nu \lambda} = {^4R}^{\sigma}_{\mu \nu \lambda} + \epsilon (K_{\mu \lambda}K^{\sigma}_{\nu}-K_{\nu \lambda}K^{\sigma}_{\mu})
\end{equation}

By contracting the indexes in the above formula we get the expression for the 4+1 scalar Riemann curvature 
\begin{equation}
\label{5Riemann}
{^5R} = {^4R} + \epsilon( K^2 - K_{\mu \nu} K^{\mu \nu} )
\end{equation}

\subsection{5-Dimensional ADM Hamiltonian}
\label{subsec:5dADMHamiltonian}

The Hilbert-Einstein  5-dimensional vacuum space action with the metric as above 

\begin{equation}
S = \int dy \;L=  \frac{1}{(16 \pi \tilde{G})} \int dy \, dx^4 \sqrt{\tilde{g}} \;\;^5{R}
\end{equation}

By using ($\ref{5Riemann}$) we can now write the action via 4+1 decomposition:

\begin{equation}
\label{admaction}
S = \frac{1}{(16 \pi \tilde{G})} \int dy d^4x \; \sqrt{\epsilon} \phi \sqrt{-g}\, ({^4R} + \epsilon( K^2 - K_{\mu \nu} K^{\mu \nu}))
\end{equation}

we used the $\sqrt{\tilde{g}} = \sqrt{\epsilon} \phi \sqrt{-g}$,  and  $ \tilde{G}$ is a 5-dimensional gravitational constant. 

Since we are considering the case of the comoving coordinates $\vec{\phi} = 0$, i.e only lapse and no shift, the extrinsic curvature is

\begin{equation}
\label{5dk}
K_{\mu \nu} = \frac{\epsilon}{2\phi} \frac{\partial g_{\mu \nu}}{\partial y},  \quad K_{A 4} = 0
\end{equation}

From ($\ref{admaction}$) and ($\ref{5dk}$) we obtain the expression for the momentum \\

\begin{equation}
\label{momentum}
\pi^{\mu \nu} = \frac{\partial L }{\partial {\overset{\star}{g}}_{\mu\nu}} = \frac{\sqrt{\epsilon}\sqrt{-g}}{16\pi \tilde{G}}\epsilon(g^{\mu \nu } K-K^{\mu \nu} ), \quad \pi_{A 4} = 0
\end{equation}

where we used $ {\overset{\star}{g}}_{\mu\nu} $ to denote the derivative with respect to the fifth coordinate $\frac{\partial g_{\mu \nu}}{\partial y}$ to  distinguish it from the usual dot symbol, the derivative with respect to time variable $x_0$

From the momentum expression and Legendre transform we obtain the Hamiltonian:

\begin{equation}
\label{HamiltonianADM}
H_{\mbox{ADM}} =  \frac{\sqrt{\epsilon}\sqrt{-g}}{16\pi \tilde{G}} \int d^4x \; \phi (\epsilon(K^2 - K^{\mu \nu}K_{\mu \nu}) - {^4R}) =  \frac{\sqrt{\epsilon}\sqrt{-g}}{16\pi \tilde{G}} \int d^4x \; \phi H
\end{equation}

Thus the Hamiltonian constraint becomes:
\begin{equation}
\label{HamiltonianConstraint}
H = {\left (\frac {\sqrt{\epsilon}\sqrt{-g}}{16 \pi \tilde{G}} \right )}(\epsilon(K^2 - K^{\mu \nu}K_{\mu \nu}) - {^4R})
\end{equation}

\subsection{4-Dimensional Supermetric }
\label{subsec:4dsupermetric}

We can now see if the 4-dimensional supermetric can be defined similar to DeWitt's $\cite{DeWitt}$  3-dimensional one. Fortunately it works. One can check that 

\begin{equation}
\label{4dsupermetric1}
2 ( K_{\mu \nu} K^{\mu \nu} - K^2 )= (K^{\mu \nu} - g^{\mu \nu} K) (K^{\alpha \beta} - g^{\alpha \beta}K) (g_{\mu \alpha} g_{\nu \beta} + g_{\mu \beta} g_{\nu \alpha} - \frac{2}{3}g_{\mu \nu} g_{\alpha \beta})
\end{equation}

or by introducing the 4-dimensional supermetric notation
\begin{equation}
G_{\mu \nu \alpha \beta} =(g_{\mu \alpha} g_{\nu \beta} + g_{\mu \beta} g_{\nu \alpha} -\frac{2}{3} g_{\mu \nu} g_{\alpha \beta})
\end{equation}

and recalling the expression ($\ref{momentum}$) for the momentum $\pi_{\mu\nu}$ we can rewrite ($\ref{4dsupermetric1}$)  as

\begin{equation}
\label{4dsupermetric2}
2 ( K_{\mu \nu} K^{\mu \nu} - K^2 )= {\left (\frac{16 \pi \tilde{G}}  {\sqrt{\epsilon}\sqrt{-g}} \right )}^2 \frac{1}{{\epsilon}^2}{\pi}^{\mu \nu} {\pi}^{\alpha \beta} G_{\mu \nu \alpha \beta}
\end{equation}

We can now obtain the expression for the ADM Hamiltonian. From ($\ref{HamiltonianConstraint}$) and ($\ref{4dsupermetric2}$)

\begin{equation}
H = -{\left (\frac{8 \pi \tilde{G}{\sqrt{\epsilon}}} {\sqrt{-g}\epsilon^2} \right )} {\pi}^{\mu \nu} {\pi}^{\alpha \beta} G_{\mu \nu \alpha \beta} - {\left (\frac {\sqrt{\epsilon}\sqrt{-g}}{16 \pi \tilde{G}} \right )}{^4R}
\end{equation}

\section[test]{Kaluza-Klein Non-Compactified Theory}
\label{sec:Kaluza-Kliein}
In this section by using Kaluza-Klein non-compactified theory, we will obtain a different expression for the action of the same physical space. It will be expressed in the terms of the massless scalar field.  By comparing it with the action in terms of extrinsic curvature obtained in the previous section we will derive  4-dimensional supermetric analog of the Wheeler-DeWitt equation.
 
We begin with the introduction to the Kaluza-Klein theory. One considers the 5-dimensional Hilbert-Einstein action  $\cite{PWessonBook}$ in vacuum 
\begin{equation}
\label{KaluzaAction}
S = \frac{1}{(16 \pi \tilde{G})} \int dy \, dx^4 \sqrt{\tilde{g}}\; \;{^5R}
\end{equation}

and by varying it with respect to the following 5-dimensional metric \\[2ex] 
$
\begin{pmatrix}
g_{\mu \nu} + {\phi}^2 A_{\mu} A_{\nu}&  \cdots  & {\phi}^2 A_{\mu} \\
\vdots  & \ddots & \vdots \\
{\phi}^2 A_{\nu} &  \cdots  &  \epsilon{\phi}^2 \\
\end{pmatrix}
$\\[2ex]

 one obtains the Einstein equation in 4-dimensional space with the electromagnetic and matter stress energy tensor on the r.h.s plus the Maxwell equation for the electromagnetic field. We consider the case $A_{\mu} = 0$ with no electromagnetic field and only the massless scalar field $\phi$ being present.  The metric then becomes exactly as considered in the previous section. The action ($\ref{KaluzaAction}$) is also the same as the one considered in the 5-dimensional ADM model above. Thus we have two formalisms describing the same physical system: 5D ADM and 5D Kaluza-Klein. \\

The 5-dimensional Riemann scalar can be expressed by using 4-dimensional Riemann scalar and the massless scalar field $\phi$ as follows:
\begin{equation}
\label{5R}
{^5R} = {^4R} + {\tilde{g}}^{44} \; {^5R}_{44} = {^4R} + \frac{\epsilon \;{^5R}_{44} }{{\phi}^2}
\end{equation}

The last term can be expressed via the 5-dimensional metric $\cite{WessonPoncedeLeon}$ to obtain
\begin{equation}
\label{5R44}
{^5R}_{44} = -\epsilon \phi \; \square \; \phi - \frac{\overset{\star}{g}^{\alpha \beta}  {\overset{\star}{g}_{\alpha \beta}}}{2} - \frac{g^{\alpha \beta} \; {\overset{\star \star}{g}_{\alpha \beta}}}{2}  + \frac{\overset{\star}{\phi}{g}^{\alpha \beta}  {\overset{\star}{g}_{\alpha \beta}}}{2 \phi} - \frac{g^{\mu \beta} g^{\alpha \nu} \overset{\star}{g}_{\alpha \beta}  {\overset{\star}{g}_{\mu \nu }}}{4}
\end{equation}

,where $\square \; \phi = g^{\mu \nu} \nabla_{\nu} \frac{\partial \phi}{\partial x^{\mu}}$

We can now rewrite the action expression ($\ref{KaluzaAction}$) by using  ($\ref{5R}$)

\begin{equation}
\label{KaluzaKleinAction}
S = \frac{1}{(16 \pi \tilde{G})} \int dy \, dx^4 \sqrt{\epsilon}\phi \; \sqrt{-g}\; \left ({^4R} +  \frac{\epsilon \; {^5R}_{44} }{{\phi}^2} \right )
\end{equation}

or by using  ($\ref{5R44}$)
\begin{equation}
\label{KaluzaKleinActionLong}
S = \frac{1}{(16 \pi \tilde{G})} \int dy \, dx^4 \sqrt{\epsilon}\phi \; \sqrt{-g}\; \left ({^4R}  -\frac{{\epsilon}^2 \; \square \phi}{\phi} \;  - \frac{\epsilon \overset{\star}{g}^{\alpha \beta}  {\overset{\star}{g}_{\alpha \beta}}}{2{\phi}^2} - \frac{\epsilon g^{\alpha \beta} \; {\overset{\star \star}{g}_{\alpha \beta}}}{2{\phi}^2}  + \frac{\overset{\star}{\phi}\epsilon {g}^{\alpha \beta}  {\overset{\star}{g}_{\alpha \beta}}}{2 {\phi}^3} - \frac{\epsilon g^{\mu \beta} g^{\alpha \nu} \overset{\star}{g}_{\alpha \beta}  {\overset{\star}{g}_{\mu \nu }}}{4{\phi}^2} \right )
\end{equation}

From ($\ref{KaluzaKleinActionLong}$) we obtain the momentum

\begin{equation}
\label{momentumphi}
{\pi}^{\phi} = \frac{\partial L}{\partial \overset{\star}{\phi}} = \frac{\sqrt{-g}}{(16 \pi \tilde{G})} \frac{\sqrt{\epsilon}\epsilon {g}^{\alpha \beta}  {\overset{\star}{g}_{\alpha \beta}}}{2 {\phi}^2}
\end{equation}


\section[test]{Wheeler-DeWitt Equation for 4-Dimensional Supermetric}
\label{sec:Wheeler-DeWitt}

By comparing the ADM and Kaluza-Klein actions  ($\ref{admaction}$) and  ($\ref{KaluzaKleinAction}$) we obtain the following equation expressing the extrinsic curvature via the massless scalar field:

\begin{equation}
\epsilon( K^2 - K_{\mu \nu} K^{\mu \nu} )=  \frac{ \epsilon \;\; {^5R}_{44} }{{\phi}^2}
\end{equation}

by using the expression ($\ref{4dsupermetric2}$) for 4-dimensional supermetric, it becomes:

\begin{equation}
\label{WDW1}
-\frac{1}{2}{\left (\frac{16 \pi \tilde{G}} {\sqrt{-g}} \right )}^2 {\pi}^{\mu \nu} {\pi}^{\alpha \beta} G_{\mu \nu \alpha \beta} =  \frac{{^5R}_{44} }{{\phi}^2}
\end{equation}

by substituting for ${^5R_{44}}$ its expression from  ($\ref{5R44}$) , we obtain:

\begin{equation}
-\frac{1}{2}{\left (\frac{16 \pi \tilde{G}} {\sqrt{-g}} \right )}^2 {\pi}^{\mu \nu} {\pi}^{\alpha \beta} G_{\mu \nu \alpha \beta} =  - \frac{\epsilon \square \; \phi}{\phi}   - \frac{\overset{\star}{g}^{\alpha \beta}  {\overset{\star}{g}_{\alpha \beta}}}{2 {\phi}^2} - \frac{g^{\alpha \beta} \; {\overset{\star \star}{g}_{\alpha \beta}}}{2 {\phi}^2}  + \frac{\overset{\star}{\phi}{g}^{\alpha \beta}  {\overset{\star}{g}_{\alpha \beta}}}{2 {\phi}^3} - \frac{g^{\mu \beta} g^{\alpha \nu} \overset{\star}{g}_{\alpha \beta}  {\overset{\star}{g}_{\mu \nu }}}{4 {\phi}^2}
\end{equation}

After substituting ${\pi}^{\phi}$ from ($\ref{momentumphi}$) 

\begin{equation}
-\frac{1}{2}{\left (\frac{16 \pi \tilde{G}} {\sqrt{-g}} \right )}^2 {\pi}^{\mu \nu} {\pi}^{\alpha \beta} G_{\mu \nu \alpha \beta} =   \frac{16 \pi \tilde{G}}{\sqrt{\epsilon} \sqrt{-g}}  \frac{1}{ \phi}{ \pi}^{\phi}{ \overset{\star}{\phi}}    - \frac{\epsilon \square \; \phi}{\phi}   - \frac{\overset{\star}{g}^{\alpha \beta}  {\overset{\star}{g}_{\alpha \beta}}}{2 {\phi}^2} - \frac{g^{\alpha \beta} \; {\overset{\star \star}{g}_{\alpha \beta}}}{2 {\phi}^2} - \frac{g^{\mu \beta} g^{\alpha \nu} \overset{\star}{g}_{\alpha \beta}  {\overset{\star}{g}_{\mu \nu }}}{4 {\phi}^2}
\end{equation}

We carry out the quantization as in DeWitt's paper  $\cite{DeWitt}$, i.e, by replacing the momenta with the quantum operators:\\[2ex]

$ {\pi}^{\mu \nu} \rightarrow  -i \frac{\partial}{\partial g_{\mu \nu}} $\\[2ex]

$ {\pi}^{\phi} \rightarrow  -i \frac{\partial}{\partial \phi} $\\

While the spacelike case $(\epsilon = 1)$ leaves us with the complex $i$ in the r.h.s of the equation, for the timelike case $\epsilon = -1$ it becomes real and we obtain the analog of the quantum Wheeler-DeWitt equation for the 4-dimensional supermetric:

\begin{multline}
\frac{1}{2}{\left (\frac{16 \pi \tilde{G}} {\sqrt{-g}} \right )}^2  G_{\mu \nu \alpha \beta}  \frac{\partial}{\partial g_{\mu \nu}} \frac{\partial}{\partial g_{\alpha \beta}} \Psi =  \pm \frac{16 \pi \tilde{G}}{ \sqrt{-g}}  \frac{1}{ \phi} { \overset{\star}{\phi}} \frac{\partial}{\partial \phi}\Psi   - \left ( \frac{\epsilon \square \; \phi}{\phi}   - \frac{\overset{\star}{g}^{\alpha \beta}  {\overset{\star}{g}_{\alpha \beta}}}{2 {\phi}^2} - \frac{g^{\alpha \beta} \; {\overset{\star \star}{g}_{\alpha \beta}}}{2 {\phi}^2} - \frac{g^{\mu \beta} g^{\alpha \nu} \overset{\star}{g}_{\alpha \beta}  {\overset{\star}{g}_{\mu \nu }}}{4 {\phi}^2} \right ) \Psi
\end{multline}

or, when writing the wave function $\Psi$ with its argument:
\begin{multline}
\frac{1}{2}{\left (\frac{16 \pi \tilde{G}} {\sqrt{-g}} \right )}^2  G_{\mu \nu \alpha \beta}  \frac{\partial}{\partial g_{\mu \nu}} \frac{\partial}{\partial g_{\alpha \beta}} \Psi({g_{\mu\nu}, \phi})  =  \pm \frac{16 \pi \tilde{G}}{ \sqrt{-g}}  \frac{1}{ \phi} { \overset{\star}{\phi}} \frac{\partial}{\partial \phi}\Psi({g_{\mu\nu}, \phi})  \\
 - \left ( \frac{\epsilon \square \; \phi}{\phi}   - \frac{\overset{\star}{g}^{\alpha \beta}  {\overset{\star}{g}_{\alpha \beta}}}{2 {\phi}^2} - \frac{g^{\alpha \beta} \; {\overset{\star \star}{g}_{\alpha \beta}}}{2 {\phi}^2} - \frac{g^{\mu \beta} g^{\alpha \nu} \overset{\star}{g}_{\alpha \beta}  {\overset{\star}{g}_{\mu \nu }}}{4 {\phi}^2} \right ) \Psi({g_{\mu\nu}, \phi})
\end{multline}\\[2ex]

We have to mention that when one uses the scalar field as internal time, the formalism depends on the time function $\phi$ and might fail if it has critical points or its gradient becomes spacelike. The obtained Wheeler-DeWitt equation for the 4-dimensional supermetric  resembles the conventional Schrödinger equation as it contains the internal time derivative in the right hand side and, unlike the original Wheeler-DeWitt equation, presents the evolution with respect to internal time.

\section{Discussion}
\label{sec:Discussion}

We considered the 4-dimensional Lorentz manifold embedded into the 5-dimensional space with the massless scalar field to be the fifth coordinate. We used the massless scalar fields as internal time and performed the ADM split by the hyper-surfaces $\Sigma_{\phi = const}$. We considered both spacelike and timelike cases. As a result we obtained the Hamiltonian formalism for the 4-dimensional supermetric similar to the DeWitt's 3-dimensional supermetric $\cite{DeWitt}$. We then studied the same physical space from the Kaluza-Klein point of view and obtained the different expression for the system action expressed via the massless scalar field. By using both actions from ADM and Kaluza-Klein we obtained the 4-dimensional analog of the Wheeler-DeWitt equation with the 4-dimensional supermetric. Finally we were able to quantize that equation similar to DeWitt's approach in $\cite{DeWitt}$, i.e, by replacing the momentum functions with the momentum operators, and obtained the quantum Wheeler-DeWitt equation for the timelike case. The obtained equation resembles the conventional Schrödinger equation as it contains the internal time derivative in the right hand side and, unlike the original WDW, presents the evolution with respect to it.

\end{document}